\documentclass[journal]{IEEEtran}

\usepackage{hyperref}
\usepackage{url}
\usepackage{graphicx}
\usepackage{algorithm}
\usepackage{algorithmic}
 \usepackage{booktabs}
 \usepackage{tabularray}
 \usepackage{mwe}
 \usepackage{amsfonts}
 \usepackage{amsmath}
 \usepackage{array}
 \usepackage{eurosym}
 \usepackage{enumitem}
 \usepackage{textcomp}
\usepackage{stfloats}
\usepackage{url}
\usepackage{verbatim}
\usepackage{graphicx}
\usepackage{subcaption}
\usepackage[font=small,labelfont=bf]{caption}
\usepackage{amssymb}
\usepackage{bbm}
\usepackage{bm}
\usepackage{tabularx}

\usepackage{hyperref}
\hypersetup{
    colorlinks = true,
    citecolor = blue,
    linkcolor = blue,
    urlcolor = blue,
    bookmarksopen = true,
    filecolor=blue
}

\usepackage{booktabs}
\usepackage{multirow}

\newcommand{\ch}{\text{ch}}
\newcommand{\dis}{\text{dis}}

\newcommand{\ev}{\text{EV}}

\let\eqref\relax
\newcommand{\eqref}[1]{(\ref{#1})}

\begin{document}

\title{Physics-Informed Reinforcement Learning for Large-Scale EV Smart Charging Considering Distribution Network Voltage Constraints}

\author{Stavros Orfanoudakis,~\IEEEmembership{Student Member,~IEEE,}
        Frans A. Oliehoek,
        Peter~Palensky,~\IEEEmembership{Senior Member,~IEEE,}
        and~Pedro~P.~Vergara,~\IEEEmembership{Senior Member,~IEEE}
\thanks{This work is funded by the HORIZON Europe Drive2X Project 101056934.}
\thanks{Stavros Orfanoudakis, Peter Palensky, and Pedro P. Vergara are with the Intelligent Electrical Power Grids (IEPG) Section, Delft University of Technology, Delft, The Netherlands (emails: {s.orfanoudakis, p.palensky, p.p.vergarabarrios}@tudelft.nl).}
\thanks{Frans Oliehoek is with the Sequential Decision Making group, Delft University of Technology, Delft, The Netherlands (email: f.a.oliehoek@tudelft.nl).}
}

\markboth{ }%
{\MakeLowercase{\textit{Orfanoudakis et al.}}}

\maketitle

\begin{abstract}
Electric Vehicles (EVs) offer substantial flexibility for grid services, yet large-scale, uncoordinated charging can threaten voltage stability in distribution networks. Existing Reinforcement Learning (RL) approaches for smart charging often disregard physical grid constraints or have limited performance for complex large-scale tasks, limiting their scalability and real-world applicability. This paper introduces a physics-informed (PI) RL algorithm that integrates a differentiable power flow model and voltage-based reward design into the Twin Delayed Deep Deterministic Policy Gradient (TD3) algorithm, enabling EVs to deliver real-time voltage support while meeting user demands. The resulting PI-TD3 algorithm achieves faster convergence, improved sample efficiency, and reliable voltage magnitude regulation under uncertain and overloaded conditions. Benchmarks on the IEEE 34-bus and 123-bus networks show that the proposed PI-TD3 outperforms both model-free RL and optimization-based baselines in grid constraint management, user satisfaction, and economic metrics, even as the system scales to hundreds of EVs. These advances enable robust, scalable, and practical EV charging strategies that enhance grid resilience and support distribution networks operation.
\end{abstract}


%
\IEEEpeerreviewmaketitle

\section{Introduction}

\IEEEPARstart{T}{he} growing penetration of electric vehicles (EVs) into distribution networks introduces significant challenges and opportunities for grid voltage regulation~\cite{STIASNY2021106696}. Uncoordinated EV charging can intensify voltage violations, posing risks to grid stability, particularly during peak demand periods~\cite{HASAN2023101216}. Meanwhile, coordinated EV charging strategies transform EVs into controllable distributed energy resources capable of alleviating these issues, especially by leveraging vehicle-to-grid (V2G) operation~\cite{mattos2024}. This dual potential highlights the crucial role of well-managed EV charging schemes, which can simultaneously maintain grid stability and facilitate the integration of renewable energy resources.

Early studies on voltage regulation via EV coordination typically employed heuristic methods~\cite{KNEZOVIC2016274} or stochastic mathematical optimization techniques.
Examples include artificial bee colony optimization~\cite{M20251095} and model predictive control (MPC), which schedule EV charging by accounting for uncertainties in renewable generation and load conditions~\cite{ELALFY2025100872}. Although MPC methods leverage forecasts to maintain voltage and frequency stability proactively, their effectiveness is often limited by uncertainty quantification inaccuracies~\cite{wevj16060292} and computational complexity~\cite{10138878}. Additionally, pricing-based mechanisms have been proposed to incentivize EV charging behaviors beneficial for voltage support~\cite{SINGH2023100972}. However, such approaches generally involve extensive offline computations and face scalability challenges, particularly when managing large EV fleets in real-time operational conditions.

To address scalability and modeling complexity challenges, reinforcement learning (RL) methods have emerged as effective solutions capable of real-time decision-making even for complex optimization tasks~\cite{SuttonReinforcementIntroduction}. Deep model-free RL algorithms, such as Deep Q Networks, have been successfully applied in a two-layer framework to jointly optimize EV charging and Volt–VAR control~\cite{8892476}. Additionally, continuous-action algorithms like Deep Deterministic Policy Gradient (DDPG) have efficiently managed EV fleet charging while explicitly considering distribution network voltage stability~\cite{liu2023DRL} and incorporating battery degradation impacts due to vehicle-to-grid (V2G) operations~\cite{SHIBL2023494}. Safe RL approaches integrate constraints in the training process to enhance grid reliability and constraint satisfaction~\cite{fan2024safeRL}, whereas model-based RL techniques leverage learned transition dynamics to improve overall decision quality~\cite{hossain2024efficient}. Furthermore, multi-agent RL strategies enable decentralized voltage control across network buses~\cite{9805763} and facilitate EV charging optimization targeting transformer lifetime extension~\cite{9756505}. 
Despite the advances summarized in Table~\ref{tab:lit}, RL methods remain limited by high-dimensional, constrained, and stochastic decision spaces that degrade sample efficiency and reliability at scale~\cite{dulac2021challenges}, while classic optimization suffers from combinatorial explosion and nonconvex network physics, limiting tractable deployment in large EV charging systems.


\begin{table*}
\centering
\caption{Comparison of EV charging control methods under voltage and grid constraints.}
\label{tab:lit}
\begin{tabularx}{\textwidth}{@{} l l c c X l r @{}}
\toprule
Reference & Method & V2G & Grid Constraints & Comments & Grid & EV Chargers \\
\midrule
\cite{KNEZOVIC2016274}      & Droop control              & No      & Phase voltage, unbalance                 & Cuts unbalance, reactive-only            & 10-bus       & 43 \\
\cite{M20251095}            & Metaheuristic Opt.           & No      & Voltage, THD                          & Improves voltage, needs data             & IEEE 33-bus  & 10 \\
\cite{ELALFY2025100872}     & Metaheuristic Opt.           & Yes     & Frequency, voltage                    & Better stability, offline retuning       & 5-bus        & 2 \\
\cite{wevj16060292}         & MPC             & Yes     & Frequency, voltage                    & Predictive control, high complexity    &IEEE 39-bus       & 5 \\
\cite{10138878}             & MPC                                   & Yes     & Frequency (islanded)                            & Inertia-like support, charging delays    & --           & 1 \\
\cite{SINGH2023100972}      & MPC            & Yes     & Voltage, power limits                 & Lowers cost, needs aggregator/comms      & 8-bus        & 3 \\
\cite{8892476}              & RL               & No      & Voltage                              & Fast coordination, training needed  & IEEE 123-bus      & -- \\
\cite{liu2023DRL}           & RL               & Yes     & Voltage, generator limits             & Cost+voltage co-optim., tuning burden    & IEEE 33-bus  & 5 \\
\cite{SHIBL2023494}         & RL             & Yes     & Voltage, transformer limit                   & Protects grid/users, simplified env.     & IEEE 33-bus  & 1 \\
\cite{fan2024safeRL}        & Safe RL          & Yes     & Voltage constraints                              & Explicit safety, higher complexity       & IEEE 33-bus  & 4 \\
\cite{9756505}              & Multi-agent RL   & Yes & Transformer thermal                         & Reduces aging, complex training          & 1-bus        & 64 \\
\textbf{Ours} & Physics-Informed RL                  & Yes     & Voltage magnitude                 & Scalable and efficient              & IEEE 123-bus      & 500 \\
\bottomrule
\end{tabularx}
\end{table*}

Physics-informed learning methods have become prominent for enhancing machine learning robustness and accuracy by directly embedding domain-specific physical equations and constraints into training processes~\cite{karniadakis2021physics}. For instance, Physics-Informed Neural Networks (PINNs) explicitly embed domain-specific equations related to EV dynamics, such as battery state-of-charge (SoC) and power consumption~\cite{10905642, lim2024evpin}, as well as networks' power flow equations~\cite{kaseb2025adaptiveinformeddeepneural}, resulting in accurate supervised learning predictions, even with limited training data. Recent advancements have extended these techniques to complex spatiotemporal prediction tasks, such as citywide EV charging demand forecasting and dynamic pricing through physics-informed graph learning \cite{KUANG2024123059}. Similarly, graph neural networks combined with deep RL have leveraged physics-informed graph attention networks to address robust voltage control challenges under partial observability~\cite{10113230}. Moreover, other studies have integrated physics-based constraint layers into RL for transient voltage control \cite{9916278}, distributed voltage regulation using photovoltaic inverters \cite{ZHANG2024109641}, and enforcing safety constraints in action selection \cite{10418941}.
However, existing approaches do not directly embed distribution network power flow and EV battery/SoC dynamics into the learning objective and updates. Instead, they typically enforce physics via action projections, constraint layers, or penalty shaping, weakly coupling grid physics to the RL algorithm. 

To overcome the scalability and efficiency shortfalls of existing methods (see Table~\ref{tab:lit}), we propose a physics-informed RL (PI-RL) algorithm\footnote{Open-sourced code at: \url{https://github.com/StavrosOrf/EV2Gym_PI-TD3}, and \url{https://github.com/distributionnetworksTUDelft/EV2Gym_PI-TD3}} tailored to city-scale EV charging while supporting the distribution network's voltage magnitude limits.
Rather than imposing physics through penalties or action filters, the proposed PI-RL embeds the power flow formulation and battery SoC dynamics into the training rollouts and reward.
By embedding the power flow formulation via differentiable reward signals directly into the learning process, the algorithm obtains richer gradient information that accelerates convergence and improves constraint satisfaction.
In detail, the proposed physics-informed Twin Delayed DDPG (PI-TD3) RL algorithm achieves higher sample efficiency, faster convergence, and fewer voltage magnitude violations in stochastic settings. 
Extensive experiments in EV2Gym~\cite{orfanoudakis2024ev2gym} on benchmark IEEE distribution networks show that PI-TD3 scales to larger EV fleets while outperforming RL and optimization baselines, thereby enabling credible, real-time coordination of city-scale EV charging.
The primary contributions of this work can be summarized as follows:
\begin{itemize}
  \item A physics-informed formulation is introduced that differentiably embeds power flow and EV battery SoC dynamics into training rollouts and the reward, enabling better enforcement of voltage magnitude limits and reducing violations without needing action filters.
  \item By embedding physics equations in the RL training process, the proposed PI-TD3 attains higher sample efficiency and faster convergence under stochastic demand, prices, and EV arrivals, compared to classic RL.
  \item The physics-informed design scales in practice enabling PI-TD3 to coordinate hundreds of chargers, supporting city-wide operation and overcoming the scalability limits of classic RL and optimization baselines.
\end{itemize}

\section{The Optimal EV Charging Problem}

In this section, the optimal EV charging problem is formalized as both a mixed-integer nonlinear programming (MINLP) problem and a Markov decision process (MDP). These formulations capture the objectives of smart EV charging while limiting voltage magnitude violations.

\subsection{Mathematical Programming Formulation} \label{app_1}

\begin{figure}[t]
  \centering  
  \includegraphics[width=\linewidth]{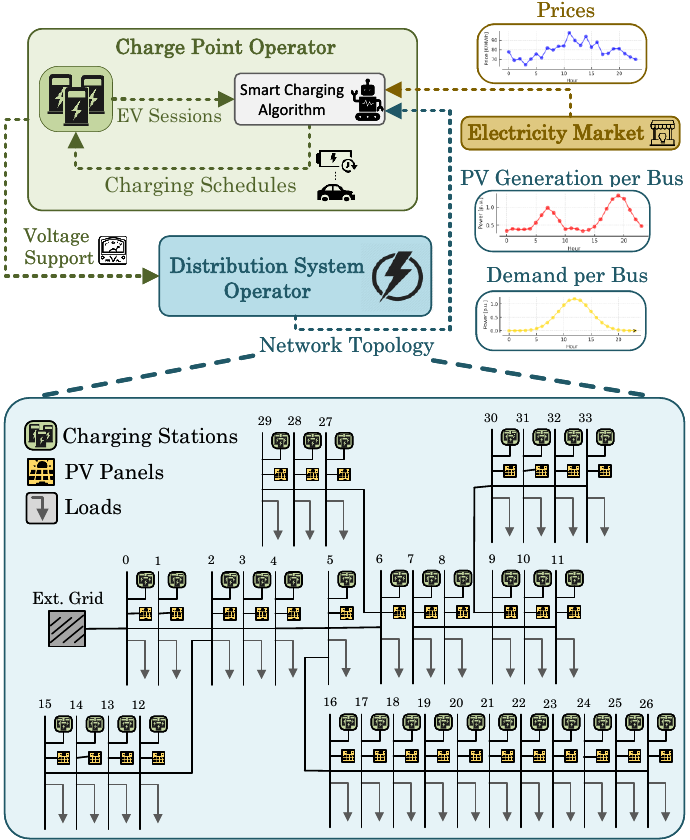} 
  \caption{Overview of the proposed problem setting illustrating an example distribution network with V2G charging stations, PV generation, and dynamic demand. The DSO provides real-time grid topology, demand, and PV generation data, while the CPO coordinates the charging of hundreds of EVs based on energy prices, respecting user constraints and enabling distribution network voltage support.}
  \label{fig:overview}
\end{figure}

The optimal EV charging problem investigated in this work is formulated on a distribution network consisting of $N$ buses, with the network topology described by the bus admittance matrix $\mathbf{Y} \in \mathbb{C}^{N \times N}$, as illustrated in Figure~\ref{fig:overview}. The problem is simulated over a discrete time horizon of $T$ steps, $t \in \mathcal{T} = \{1, \ldots, T\}$. At each time step, the Charge Point Operator (CPO) determines the charging and discharging power, $p^\ch_{i,t}$ and $p^\dis_{i,t}$, for each charging station $i \in \mathcal{I}$.
Charging stations are geographically distributed and grouped according to the buses to which they are connected, indexed by $n \in \mathcal{N} = \{1, \ldots, N\}$. For each bus $n \in \mathcal{N}$, let $\mathcal{I}_n \subset \mathcal{I}$ denote the set of charging stations associated with that bus. The distribution network is described by the admittance vector $\mathbf{Z} \in \mathbb{C}^{N}$ and the reduced admittance matrix $\mathbf{L} \in \mathbb{C}^{N \times N}$.

During operation, the Distribution System Operator (DSO) provides the CPO with real-time information including, active and reactive demands $p^L_{n,t}$ and $q^L_{n,t}$, as well as photovoltaic (PV) generation $p^{\textit{PV}}_{n,t}$ at every bus $n \in \mathcal{N}$. Although the CPO does not have access to forecasts of future EV arrivals or network states, each EV, upon arrival at charging station $i$, communicates its expected departure time $t^{d}_{i}$ and desired battery capacity at departure $e^*_{i}$. The real-time battery energy $e_{i,t}$ for every connected EV is assumed known, as is standard in V2G-enabled charging communication protocols~\cite{OCPP2.1}.

Within this setting, the CPO seeks to maximize profit, satisfy user charging needs, while minimizing voltage magnitude violations and enforcing battery operational constraints. The expression in \eqref{eq:opt3} defines the optimization objective: the first term penalizes voltage deviations at each bus, the second term accounts for net profit from charging and discharging, and the third term penalizes deviations from user-specified energy requirements. Each term is weighted ($\lambda_1,
\lambda_2$, and $\lambda_3$) to balance the priority of each objective.
Therefore, the overall EV charging optimization problem is defined as:
\begin{align}
&\min_{p^\ch,\,p^\dis}\sum_{t\in\mathcal{T}}\Biggl\{
  \lambda_1 \sum_{n\in\mathcal{N}}
  \min\Bigl[0,\;0.05 - \bigl|1 - V(p_{i,t}^\ch,p_{i,t}^\dis)_{n,t}\bigr|\Bigr]
\notag \\[2pt]
&\qquad + \sum_{i\in\mathcal{I}}\Bigl[
     \lambda_2 \Delta t\,\bigl(\Pi^\ch_t\,p^\ch_{i,t}\,\omega^\ch_{i,t}
     - \Pi^\dis_t\,p^\dis_{i,t}\,\omega^\dis_{i,t}\bigr)
\label{eq:opt3} \\   
&\qquad + \lambda_3\sum_{j\in\mathcal{J}_i}
    \Bigl(\sum_{s=t^a_{j,i}}^{t^d_{j,i}}
      \bigl(p^\ch_{i,s}\,\omega^\ch_{i,s}-p^\dis_{i,s}\,\omega^\dis_{i,s}\bigr)
      - e^*_{j,i}\Bigr)^2
  \Bigr]
\Biggr\}
\notag
\end{align}
Subject to:
\begin{flalign}
& p^{\ev}_{n,t} = \sum_{i\in \mathcal{I}_n} \big(p^\ch_{i,t}\,\omega^\ch_{i,t} - p^\dis_{i,t}\,\omega^\dis_{i,t}\big)
  & \forall n \in \mathcal{N},\; \forall t \in \mathcal{T} &
    \label{eq:opt3.0}
\end{flalign}
\vspace{-3mm}
\begin{flalign}
& s_{n,t} = (p^{\textit{L}}_{n,t} + p^{\textit{PV}}_{n,t} + p^{\ev}_{n,t}) + q^L_{n,t}  j
  & \forall n \in \mathcal{N},\; \forall t \in \mathcal{T} &
    \label{eq:opt3.001}
\end{flalign}
\vspace{-3mm}
\begin{flalign}
& v^{(0)}_{n,t} = 1 + 0j
  && \forall n \in \mathcal{N},\; \forall t \in \mathcal{T},
  \label{eq:opt3.002}
\end{flalign}
\vspace{-5mm}
\begin{flalign}
& v_{n,t}^{(\kappa+1)} = Z_{n} + \sum_{n'\in \mathcal{N}} L_{n n'}\, \overline{\left(\frac{s_{n',t}}{v_{n',t}^{(\kappa)}}\right)}
  && \forall \kappa,\, n \in \mathcal{N},\, t \in \mathcal{T}
  \label{eq:opt3.003}
\end{flalign}
\vspace{-3mm}
\begin{flalign}
& V(p^\ch,p^\dis)_{n,t} = v^{(K)}_{n,t}
  && \forall n \in \mathcal{N},\, t \in \mathcal{T},
  \label{eq:opt3.004}
\end{flalign}
\vspace{-3mm}
\begin{flalign}
& \underline{e}_{i} \leq e_{i,t} \leq \overline{e}_{i} 
  &  \forall i \in \mathcal{I}, \ \forall t \in \mathcal{T} &
    \label{eq:opt3.3} 
\end{flalign}
\vspace{-3mm}
\begin{flalign}
& e_{i,t} = e_{i,t-1} + (p^\ch_{i,t}\,\omega^\ch_{i,t} + p^\dis_{i,t}\,\omega^\dis_{i,t}) \cdot \Delta t 
  &  \forall i \in \mathcal{I}, \forall t \in \mathcal{T} &
\label{eq:opt3.4}
\end{flalign}
\vspace{-3mm}
\begin{flalign}
& e_{i,t} = e^{a}_{i} \qquad \text{if}~ t = t^a_{i}
  &  \forall i \in \mathcal{I}, \ \forall t \in \mathcal{T} &
\label{eq:opt3.5}
\end{flalign}
\vspace{-3mm}
\begin{flalign}
& \underline{p}^{\ch}_{i}\leq p^\ch_{i,t} \leq \overline{p}^{\ch}_{i} 
  &  \forall i \in \mathcal{I}, \ \forall t \in \mathcal{T} &
\label{eq:opt3.6}
\end{flalign}
\vspace{-3mm}
\begin{flalign}
& \underline{p}^{\dis}_{i} \leq p^\dis_{i,t} \leq \overline{p}^{\dis}_{i}
  &  \forall i \in \mathcal{I}, \ \forall t \in \mathcal{T} &
\label{eq:opt3.7}
\end{flalign}
\vspace{-3mm}
\begin{flalign}
& \omega^\ch_{i,t} + \omega^\dis_{i,t} \leq 1
  &  \forall i \in \mathcal{I}, \ \forall t \in \mathcal{T} &
 \label{eq:opt3.11}
 \end{flalign}
In the objective function \eqref{eq:opt3}, the decision variables $p_{i,t}^\ch$ and $p_{i,t}^\dis$ define the charging and discharging power of each EV, while $\Pi_t^\ch$ and $\Pi_t^\dis$ designate the electricity price per kWh.
Constraint~\eqref{eq:opt3.0} defines the total EV charging and discharging power $p^{\ev}_{n,t}$ injected at bus $n$ and time $t$, as the sum over all charging stations $i \in \mathcal{I}_n$, where $p^\ch_{i,t}$ and $p^\dis_{i,t}$ denote the charging and discharging power, and $\omega^\ch_{i,t},\,\omega^\dis_{i,t}$ are their respective binary activation variables. The total complex power injection $s_{n,t}$ at each bus \eqref{eq:opt3.001} aggregates active load $p^{\textit{L}}_{n,t}$, PV generation $p^{\textit{PV}}_{n,t}$, EV charging power $p^{\ev}_{n,t}$, and reactive load $q^L_{n,t}$.
The voltage magnitude $v_{n,t}$ at each bus is computed iteratively~\cite{GIRALDO2022108326}: initialization is set by \eqref{eq:opt3.002}; \eqref{eq:opt3.003} updates the voltage using the reduced grid admittance parameters $\mathbf{Z}$ and $\mathbf{L}$, the total complex power $s_{n,t}$, and the previous voltage iterate; the process repeats for $\kappa$ iterations, yielding the final voltage magnitude $V(p^\ch,p^\dis)_{n,t}$ in \eqref{eq:opt3.004}.
EVs' battery dynamics are enforced by~\eqref{eq:opt3.3}--\eqref{eq:opt3.5}: $e_{i,t}$ denotes the battery energy of the  EV parked at charger $i$ at time $t$, bounded by minimum and maximum values $\underline{e}_i$ and $\overline{e}_i$. The state is updated each timestep according to charging/discharging actions \eqref{eq:opt3.4} and initialized to $e^a_{i}$ at arrival time $t^a_{i}$ \eqref{eq:opt3.5}.
Charging and discharging power limits, $\underline{p}^\ch_i$, $\overline{p}^\ch_i$, $\underline{p}^\dis_i$, and $\overline{p}^\dis_i$, are imposed by~\eqref{eq:opt3.6} and~\eqref{eq:opt3.7}. Finally, \eqref{eq:opt3.11} ensures that a charger cannot charge and discharge simultaneously.

\subsection{Markov Decision Processes for EV Charging }

The optimal EV charging problem can also be formulated as an MDP $(\mathcal{S}, \mathcal{A}, \mathcal{P}, \mathcal{R})$. At each time step $t \in \mathcal{T}$, the state vector $\mathbf{s}_t$ is given by
\begin{equation}
  \mathbf{s}_t = \Big[    
    \sin(\text{h}_t),
    \cos(\text{h}_t),
    \Pi^\ch_t,\
    \mathbf{p}_t,
    \mathbf{q}_t, 
    \mathbf{SoC}_t,
    \mathbf{t}^\text{left}_t,
    \mathbf{b}_t  
\Big],
\label{eq:state_func}
\end{equation}
where $\sin(\text{h}_t)$ and $\cos(\text{h}_t)$ represent the hour ($\text{h}$) of the day as cyclical features. The net active power ($p^{\textit{PV}}_{n,t} - p^{\textit{L}}_{n,t}$), $\mathbf{p}_t = [p_{1,t}, \ldots, p_{N,t}]$ and reactive power $\mathbf{q}_t = [q_{1,t}, \ldots, q_{N,t}]$ injections at each bus $n \in \mathcal{N}$, $\mathbf{SoC}_t = [\mathrm{SoC}_{1,t}, \ldots, \mathrm{SoC}_{|\mathcal{I}|,t}]$ contains the state-of-charge of each connected EV with $\mathrm{SoC}_{i,t} = e_{i,t}/\overline{e}_i$, $\mathbf{t}^{\mathrm{left}}_t = [t^{\mathrm{left}}_{1,t}, \ldots, t^{\mathrm{left}}_{|\mathcal{I}|,t}]$ is the vector of remaining time to departure for each EV with $t^{\mathrm{left}}_{i,t} = t^d_i - t$, $\mathbf{b}_t = [b_{1}, \ldots, b_{|\mathcal{I}|}]$ specifies the bus index to which each charger is connected, and $\Pi^\ch_t$ is the current electricity price.

The action vector at each time step is
$\mathbf{a}_t = [a_{1,t}, \ldots, a_{|\mathcal{I}|,t}]^\top \in [-1,1]^{|\mathcal{I}|}$,
where $a_{i,t}$ is the normalized charging action ($a_{i,t}>0$) or discharging ($a_{i,t}<0$) for EV $i$, with $a_{i,t}=0$ denoting no action.
The transition function $\mathcal{P}$ determines the evolution of the system based on the chosen actions, grid power flow, and other unknown system dynamics, such as EV arrivals, PV generation, and load profiles.

The reward function $R(\cdot)$ is designed to closely mirror the objective of the mathematical programming formulation in \eqref{eq:opt3}, balancing voltage magnitude regulation, energy costs, and user satisfaction. Specifically, the reward $r_t$ is defined as the outcome of the reward function:
\begin{align}
R(\mathbf{s}_t,\mathbf{a}_t) &= \lambda_1 \sum_{n\in\mathcal{N}}
        \min\bigl\{0,\;0.05 - \lvert 1 - V_{n,t}(.)\rvert\bigr\} \nonumber \\
    & \hspace*{-1em}+ \sum_{i\in\mathcal{I}}\Bigl[
        \lambda_2 \Delta t\,\bigl(\Pi^{\ch}_{t}p^{\ch}_{i,t} - \Pi^\dis_{t}p^{\dis}_{i,t}\bigr)
        + \lambda_3 \cdot \psi_{i,t}
      \Bigr],
\label{eq:reward}
\end{align}
where the first term penalizes voltage magnitude violations at each bus. Note that the voltage magnitude at bus $n$ and step $t$ is described by \eqref{eq:opt3.0}-\eqref{eq:opt3.004}, and ultimately $V_{n,t}(\mathbf{a}_t)$ is a function of charging actions..
The second term represents the net revenue from charging and discharging activities based on electricity prices $\Pi^{\ch}_t$ and $\Pi^{\dis}_t$, and $\psi_{i,t}$ is a user satisfaction term that incentivizes each EV to maintain a minimum SoC as it approaches its departure time.
Unlike the original mathematical programming objective \eqref{eq:opt3}, which is sparse and directly penalizes deviations from the total energy target upon departure, the RL reward employs a denser signal defined as:
\begin{equation}    
\psi_{i,t} = \max\big\{0,\, \mathrm{SoC}^* - \mathrm{SoC}_{i,t}\big\} \cdot \mathbb{I}[t^\mathrm{left}_{i,t} < \epsilon],
\end{equation}
where $\mathrm{SoC}_{i,t}$ is the current state-of-charge of EV $i$ at time $t$, $\mathrm{SoC}^*$ is a target minimum SoC (e.g., $90\%$), $\epsilon$ is a threshold defining the proximity to departure, and $\mathbb{I}[\cdot]$ is the indicator function. The $\psi_{i,t}$ term penalizes the agent if any EV approaches departure with insufficient SoC, thereby encouraging timely charging to meet user expectations by the time of departure, while also providing a dense training signal. Here, as in the expression \eqref{eq:opt3}, the coefficients $\lambda_1$, $\lambda_2$, and $\lambda_3$ are chosen to match the weighting of the respective terms in the original MINLP formulation.
This consistency ensures that the physics-informed RL agent optimizes towards the same operational goals as the mathematical programming approach.

\subsection{RL for EV Charging}

RL can solve sequential decision-making problems expressed as MDPs, such as the EV charging problem described above, by learning a policy $\pi$ that maps the observed state of the system to charging or discharging actions~\cite{SuttonReinforcementIntroduction}. The agent’s objective is to maximize the expected cumulative reward, mathematically expressed as:
\begin{equation}
J(\pi) = \mathbb{E}_\pi\left[\sum_{t=0}^\infty \gamma^t r_t\right],
\label{eq:010}
\end{equation}
where $r_t$ denotes the reward at time $t$, $\gamma$ is a discount factor, and the expectation is taken over the stochastic evolution of the environment under the policy $\pi$. 
Furthermore, the state-action value function, or Q-function, is central to RL representing the expected cumulative reward obtained by taking action $\mathbf{a}_t$ in state $\mathbf{s}_t$ and subsequently following a policy $\pi$. The Q-function is recursively defined by the Bellman equation:
\begin{equation}
Q(\mathbf{s}_t, \mathbf{a}_t) = r_t + \gamma\, \mathbb{E}_{\mathbf{s}_{t+1} \sim P(\cdot|\mathbf{s}_t, \mathbf{a}_t)}\big[ Q(\mathbf{s}_{t+1}, \pi(\mathbf{s}_{t+1})) \big],
\label{eq:bellman_exp}
\end{equation}
where $r_t$ denotes the immediate reward, and $P(\mathbf{s}_{t+1}|\mathbf{s}_t,\mathbf{a}_t)$ is the transition probability between states. This recursive relationship links current and future value estimates, allowing RL algorithms to iteratively improve their policies using only sampled transitions and rewards. As a result, near-optimal charging strategies can be learned even when the underlying system dynamics are partially known.

\section{Physics-Informed RL for EV Charging}

Unlike the standard MDP formulation presented above, which disregards distribution network constraints, the proposed framework leverages power flow formulation to facilitate efficient and grid-aware policy learning through model-based rollouts combined with gradient-based optimization.

\subsection{From Model-Free to Physics-Informed RL}

In classic model-free RL, the environment is treated as a black box, and state transitions are learned solely from sampled experience. In contrast, PI-RL leverages known, differentiable components of the system (physics), such as the SoC update for each EV. This allows for more accurate and efficient learning by directly modeling the underlying EV battery dynamics. To ensure that EV battery constraints \eqref{eq:opt3.3} and \eqref{eq:opt3.4} are satisfied at every step, the SoC transition update is implemented as a piecewise, differentiable function:
\begin{equation}
\mathrm{SoC}_{i,t+1} =
\begin{cases}
1, & \text{if } x_{i,t+1} > 1 \\[4pt]
\underline{\mathrm{SoC}}_i, & \text{if } x_{i,t+1} < \underline{\mathrm{SoC}}_i \\[4pt]
x_{i,t+1}, & \text{otherwise}
\end{cases}
\label{eq:t1}
\end{equation}
where $x_{i,t+1} = \mathrm{SoC}_{i,t} + \dfrac{\Delta t\, a_{i,t}\, \overline{p}_{i,t}}{\overline{e}_i}$, $a_{i,t}$ is the charging/discharging action  and $\underline{\mathrm{SoC}}_i$ is the minimum SoC while doing V2G discharging. 
For clarity of presentation, the charging efficiency factor has been excluded from \eqref{eq:t1}; nevertheless, it can be incorporated in a straightforward manner without necessitating any modification to the algorithm.
Thus, \eqref{eq:t1} accurately and completely describes the EV battery transition given any charging action.

\begin{figure}[t]
  \centering  
  \includegraphics[width=1\linewidth]{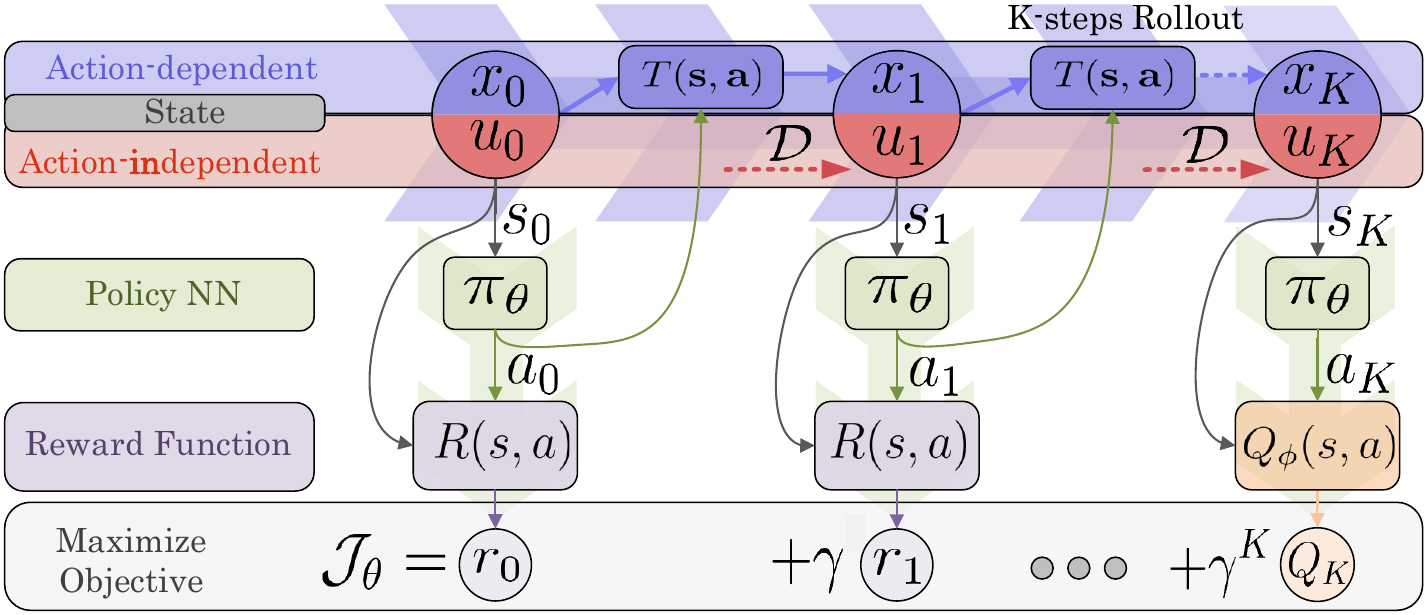} 
  \caption{
The policy network $\pi_\theta$ generates actions that, together with the known environment dynamics (e.g., SoC and voltage updates) and sampled exogenous variables (e.g., loads, prices, and EV arrivals), are used to simulate $K$-step rollouts through the differentiable transition $T(s,a)$ and reward $R(s,a)$ functions. This enables direct gradient propagation from cumulative rewards back through the rollout.
}
  \label{fig:method}
\end{figure}

Some aspects of the transition are unknown or stochastic and are independent of the actions taken, e.g., future demand at each bus, electricity price signals, and the arrival and departure of new EVs. 
These elements are difficult to model or forecast, but can be sampled from historical data. In practice during the training phase, these unknown exogenous variables are sampled from a replay buffer $\mathcal{D}$, which stores past system trajectories, so that model-based rollouts are grounded in realistic scenarios.
Therefore, as shown in Figure~\ref{fig:method}, the state $\mathbf{s}_t=\{\mathbf{x_t}, \mathbf{u}_t\}$ is defined as the combination of action-dependent variables $\mathbf{x}= \{\mathbf{SoC}_t\}$, and the action-independent variables $\mathbf{u}= \{ \sin(\text{h}_t), \cos(\text{h}_t), \Pi^\ch_t,\, \mathbf{p}_t, \mathbf{q}_t,{\mathbf{t}^\text{left}_t}, \mathbf{b}_t \}$.

The reward function $R(\mathbf{s}, \mathbf{a})$ for this problem is fully known and differentiable, as defined in~\eqref{eq:reward}. All variables required for its computation, such as voltage magnitudes, charging and discharging power, electricity prices, and user satisfaction terms, are either included in the state, sampled as exogenous variables, or, in the case of network parameters like the grid admittance matrix, remain constant and are hardcoded throughout the simulations. As a result, for any given state-action pair, the reward can be deterministically computed. 

With the transition function $T(\mathbf{s}, \mathbf{a})$ fully specified, using sampled exogenous trajectories from $\mathcal{D}$ for the unknown variables, any state in a trajectory can be recursively computed as $\mathbf{s}_{t+1} = T(\mathbf{s}_t, \mathbf{a}_t)$.
Also, given the deterministic reward function $R(\mathbf{s}, \mathbf{a})$, it becomes possible to efficiently simulate future trajectories and compute the corresponding rewards for any sequence of actions.
Using this approach, the Bellman equation can be rolled out over $K$ steps as a $K$-step expansion:
\begin{align}
\centering
Q&(\mathbf{s}_t, \mathbf{a}_t) =\; R(\mathbf{s}_t, \mathbf{a}_t) 
+ \gamma R(\mathbf{s}_{t+1}, \mathbf{a}_{t+1}) + \cdots \nonumber \\
&+ \gamma^{K-1} R(\mathbf{s}_{t+K-1}, \mathbf{a}_{t+K-1}) 
+ \gamma^K Q(\mathbf{s}_{t+K}, \mathbf{a}_{t+K}),
\label{eq:kstep}
\end{align}
where $\mathbf{s}_{t+1} = T(\mathbf{s}_t, \mathbf{a}_t)$ and $\mathbf{a}_{t+1} = \pi_\theta(\mathbf{s}_{t+1})$ is the output of the actor policy neural network $\pi$ with parameters $\theta$.

Since both $R(\mathbf{s}, \mathbf{a})$ and $T(\mathbf{s}, \mathbf{a})$ are known for the sampled trajectories, the actor network $\pi_\theta$ can be optimized by directly backpropagating gradients through these simulated rollouts. The policy gradient update can thus be computed as:
\begin{align}
\nabla_\theta J(\theta) \approx\;& \mathbb{E}_{\tau \sim \mathcal{D}} \Bigg[
\sum_{j=0}^{K-1} \gamma^j \nabla_\theta R(\mathbf{s}_{t+j}, \pi_\theta(\mathbf{s}_{t+j})) \nonumber \\
&+ \gamma^K \nabla_\theta Q_\phi(\mathbf{s}_{t+K}, \pi_\theta(\mathbf{s}_{t+K}))
\Bigg],
\label{eq:policy_grad}
\end{align}
where $\tau$ denotes a trajectory segment sampled from the replay buffer.
During training, length-$K$ trajectories $\{\mathbf{u}_{t:t+K-1}\}$ are sampled from the replay buffer $\mathcal{D}$ and treated as fixed within each rollout, so that the trajectory evolves deterministically under the known, (piecewise) differentiable transition and reward models.
This rollout process, together with the calculation of the total cumulative objective, is illustrated in Figure~\ref{fig:method}.
For example, a three-step trajectory of exogenous variables $\tau = \{\mathbf{u}_t,\mathbf{u}_{t+1},\mathbf{u}_{t+2}\}$ is sampled from $\mathcal{D}$ and held fixed. The policy $\pi_\theta$ then outputs $\{\mathbf{a}_t,\mathbf{a}_{t+1},\mathbf{a}_{t+2}\}$, the states evolve via $\mathbf{s}_{t+j+1}=T(\mathbf{s}_{t+j},\mathbf{a}_{t+j};\mathbf{u}_{t+j})$, rewards are computed deterministically, and gradients flow only through $(\pi_\theta,T,R)$ with no backpropagation through the sampled $\tau$, which is treated as constants during the rollout.
This approach provides direct and informative gradient feedback based on the ``physics'' of the system described by $R(\mathbf{s}, \mathbf{a})$ and $T(\mathbf{s}, \mathbf{a})$, thereby enabling the policy to directly learn how charging decisions influence grid voltages, SoC evolution, and long-term operational rewards.

\begin{algorithm}[t]
\caption{Physics-Informed TD3}
\begin{algorithmic}[1]
\label{alg:1}
\STATE \textbf{Initialize} critics $Q_{\phi_1}, Q_{\phi_2}$ and actor $\pi_\theta$ with parameters $\phi_1, \phi_2, \theta$
\STATE \textbf{Initialize} target networks: $\phi'_1 \gets \phi_1$, $\phi'_2 \gets \phi_2$, $\theta' \gets \theta$
\FOR{$t = 1$ to $T$}
    \STATE Select action with exploration noise: $\mathbf{a}_t \sim \pi_\theta(\mathbf{s}_t) + \epsilon, \; \epsilon \sim \mathcal{N}(0, \sigma)$
    \STATE Execute $\mathbf{a}_t$ in environment, observe reward $r_t$ and next state $\mathbf{s}_{t+1}$
    \STATE Store $(\mathbf{s}_t, \mathbf{a}_t, r_t, \mathbf{s}_{t+1})$ in replay buffer $\mathcal{D}$
    \STATE Sample mini-batch $\{\tau_i\}_{i=1}^B$ of length-$K$ trajectories from $\mathcal{D}$, $\tau_i = (\mathbf{s}_{0,i}, \mathbf{a}_{0,i}, r_{0,i}, \ldots, \mathbf{s}_{K,i}, \mathbf{a}_{K,i}, r_{K,i})$
    \STATE $\tilde{\mathbf{a}}_{1,i} \gets \pi_{\theta'}(\mathbf{s}_{1,i}) + \epsilon$, $\epsilon \gets \mathrm{clip}(\mathcal{N}(0,\sigma),-c,c)$
    \STATE $y_i \gets r_{0,i} + \gamma \min_{j \in \{1,2\}} Q_{\phi'_j}(\mathbf{s}_{0,i}, \tilde{\mathbf{a}}_{1,i})$
    \STATE \textbf{Update critics} by minimizing:
    \[
      \phi_j \gets \min_{\phi_j}
      \sum_{i=1}^B \Big(y_i - Q_{\phi_j}(\mathbf{s}_{0,i}, \mathbf{a}_{0,i})\Big)^2,\; (j = 1,2)
    \]
    \STATE \textbf{Update actor} using $\nabla_\theta J(\theta)$ from~\eqref{eq:policy_grad}
    \STATE \textbf{Update target networks}:
    \[
      \phi'_j \gets \tau \phi_j + (1-\tau)\phi'_j,\; j \in \{1,2\}
    \]
    \[
      \theta' \gets \tau \theta + (1-\tau)\theta'
    \]
\ENDFOR
\end{algorithmic}
\end{algorithm}

\subsection{Physics-Informed TD3 for EV Charging}

The proposed physics-informed formulation for EV charging can be integrated with a range of RL algorithms for continuous control~\cite{xing2024stabilizing}, such as Sof Actor Critic (SAC) and DDPG. However, TD3~\cite{pmlr-v80-fujimoto18a} was selected as the backbone due to its superior performance in the EV charging setting~\cite{Orfanoudakis2024}, where long horizons, continuous actions, and voltage magnitude violation penalties pose particular challenges. 
The proposed PI-TD3 algorithm is described in Algorithm~\ref{alg:1}.
At the start of each epoch, new transitions are collected by executing the current policy with exploration noise (lines 4–6). During each training iteration, a mini-batch of $K$-step trajectories $\{\tau_i\}_{i=1}^B$ is sampled from the replay buffer $\mathcal{D}$ (line 7). The twin critics $Q_{\phi_1}$ and $Q_{\phi_2}$ are updated using target values computed from the replayed transitions (lines 8–10), minimizing mean-squared error. The actor network $\pi_\theta$ is then updated (line 11) with the policy gradient derived from the multi-step rollout objective defined in~\eqref{eq:policy_grad}, where gradients flow through the differentiable transition and reward models. Target networks are softly updated to ensure training stability (line 12).

This physics-informed formulation enables the PI-TD3 algorithm to scale effectively to hundreds of EVs. In particular, by embedding the power flow formulation \eqref{eq:opt3.0}-\eqref{eq:opt3.004} via differentiable reward signals \eqref{eq:reward} directly into the learning process, the algorithm obtains richer gradient information that accelerates convergence and improves constraint satisfaction. The use of model-based rollouts further enhances sample efficiency, reducing reliance on environment interactions. Unlike classical optimization approaches, whose computational complexity grows exponentially with the number of decision variables, the proposed formulation leverages neural approximations of system dynamics and differentiable constraints, allowing training complexity to grow in a more tractable manner with fleet size. Compared to conventional RL methods, which lack access to such physics-guided gradients, the proposed PI-TD3 algorithm achieves more stable and scalable learning, making it suitable for real-world deployment in large urban charging networks.


\section{Experimental Results}

In this section, the performance of the proposed PI-TD3 algorithm is systematically evaluated. Average results across multiple scenarios are reported, detailed analyses of specific cases are provided, generalization to different grid loadings is assessed, and scalability is demonstrated using a substantially larger test system.

\subsection{Experimental Setup}

To assess the effectiveness of the proposed PI-TD3 algorithm, experiments were conducted on a modified IEEE 34-bus distribution network, with each bus hosting 4--5 V2G-enabled charging stations for a total of 150 charging points. EV arrivals and departures followed realistic daily and weekly patterns derived from the ElaadNL dataset. All scenario generation was performed using the EV2Gym simulator~\cite{orfanoudakis2024ev2gym}, which models EV fleet behavior, grid topology, and user sessions with high fidelity, and was enhanced with the RL-ADN~\cite{HOU2025100457} power flow module for fast, accurate voltage magnitude calculations. The simulator included detailed EV parameters based on field data, such as battery capacity, charging rates, and efficiency.
The following weights ( $\lambda_1= -5\times10^4, \lambda_2=1, \lambda_3=-10$) were used in the problem formulation (\ref{eq:opt3}) to balance the priority of each objective. 
Each scenario consisted of 300 steps, each representing 15 minutes of simulated time. In every step, the operator had to determine in real-time the charging action for all EVs. Following standard operation and safety procedures, a voltage limit of lower limit 0.95 and higher limit 1.05 was selected.
This setup ensures a realistic benchmarking environment for RL algorithms under operational scenarios representative of real-world distribution networks.

All RL experiments were conducted on the DelftBlue high-performance computing cluster~\cite{DHPC2024}. Each RL algorithm was trained independently until convergence, with training durations ranging from 5 to 10 hours for simpler scenarios, and up to 48 hours for larger grids. To ensure a fair comparison, the default hyperparameter settings recommended in the literature for each baseline were used, including learning rates, discount factors, batch sizes, and exploration noise levels. All models were implemented in PyTorch and trained using the Adam or AdamW optimizer.
Performance was averaged across multiple random seeds to assess statistical robustness, and convergence was monitored by tracking moving averages of episode returns.  

\begin{table*}[t]
\centering
\caption{Average performance of the best trained models over 50 evaluation scenarios on the IEEE 34-bus network with 150 EV chargers.}

\label{tab:mainresults}
\begin{tblr}{
  cells = {c},
  hline{1,9} = {-}{0.08em},
  hline{2} = {-}{},
}
Algorithm & {Costs \\{[}€]} & {User \\Satisfaction [\%]} & {Total V.V.\\per bus [-]} & {Total~V.V. \\per step [-]} & {Voltage Violation\\{[}p.u.]} & {Tot. Energy \\Ch.~~[MWh]} & {Tot. Energy\\Dis.~[MWh]} & {Step time\\~[sec/step]}\\
CAFAP & $-2545$ ±$1581$ & $100.0$ ±$0.0$ & $169.5$ ±$157.1$ & $36.5$ ±$29.4$ & $-0.790$ ±$0.991$ & $12.6$ ±$0.6$ & $0.00$ ±$0.00$ & $0.003$\\
No Charging & $0$ ±$0$ & $52.1$ ±$1.4$ & $106.6$ ±$123.0$ & $24.5$ ±$25.1$ & $-0.430$ ±$0.655$ & $0.0$ ±$0.0$ & $0.00$ ±$0.00$ & $0.002$\\
SAC & $-291$ ±$189$ & $57.7$ ±$1.3$ & $111.7$ ±$125.7$ & $25.6$ ±$25.5$ & $-0.458$ ±$0.679$ & $20.3$ ±$0.9$ & $18.87$ ±$0.92$ & $0.017$\\
PPO & $-905$ ±$583$ & $69.4$ ±$1.1$ & $121.4$ ±$132.4$ & $27.7$ ±$26.6$ & $-0.505$ ±$0.737$ & $6.4$ ±$0.4$ & $1.80$ ±$0.17$ & $0.007$\\
TD3 & $-1900$ ±$1267$ & $90.2$ ±$4.5$ & $127.8$ ±$134.4$ & $30.0$ ±$27.5$ & $-0.482$ ±$0.692$ & $12.6$ ±$0.6$ & $2.58$ ±$1.06$ & $0.009$\\
PI-TD3 \textbf{(Ours)} & $-2025$ ±$1314$ & $95.6$ ±$3.6$ & $104.2$ ±$126.4$ & $25.3$ ±$27.4$ & $-0.364$ ±$0.586$ & $14.3$ ±$0.7$ & $2.86$ ±$1.04$ & $0.010$\\
MPC (Oracle) & $-1640$ ±$1203$ & $99.9$ ±$0.6$ & $98.7$ ±$125.3$ & $24.3$ ±$27.8$ & $-0.321$ ±$0.545$ & $26.0$ ±$2.0$ & $13.46$ ±$1.97$ & $-$
\end{tblr}
\end{table*}

\subsection{Baseline Methods \& Evaluation Metrics}

To benchmark the performance of the proposed PI-TD3 algorithm, several representative baselines and state-of-the-art algorithms for EV smart charging were selected. These include: (i) Charge as Fast as Possible (CAFAP), a simple heuristic in which each EV is charged at maximum rate immediately upon connection; (ii) a no-charging reference (No Charging); (iii) three widely used model-free RL algorithms, Soft Actor Critic (SAC), Proximal Policy Optimization (PPO), and standard TD3; and (iv) an oracle MPC method that assumes perfect knowledge of future system states and EV demands. While the oracle MPC is not feasible in practical deployments, it provides a useful upper bound on achievable performance under ideal information. All the RL algorithms used the same state and reward formulations to have a fair comparison with the proposed PI-TD3.

All algorithms were evaluated in a deliberately overloaded network scenario, where the distribution grid operates under high load conditions and dense EV integration. This scenario was designed to rigorously test the robustness of each method, as voltage magnitude violations may occur even in the absence of active charging (No Charging baseline). The resulting environment poses a challenging benchmark for coordinating large-scale EV charging while maintaining voltage stability.

Evaluation metrics were selected to reflect the multi-objective nature of the optimization problem in~\eqref{eq:opt3}. These include total charging cost, average user satisfaction (quantified as the ratio of SoC at departure to target SoC for each EV), and three distinct voltage magnitude violation metrics: total voltage magnitude violations per bus over the evaluation, the number of steps with at least one voltage magnitude violation (Total V.V. per step), and the aggregate absolute per-unit voltage magnitude violations across all buses. Additionally, to further characterize the performance of each approach, total energy charged and discharged by the EV fleet and the average execution time per step were recorded. 

\subsection{Comparison with Baseline Algorithms}

Table~\ref{tab:mainresults} summarizes the average performance and standard deviation of all algorithms in 50 scenarios. Notably, the proposed PI-TD3 algorithm achieves a favorable balance among all operational objectives (costs, user satisfaction, and voltage violations).
PI-TD3 delivers $14.3~\mathrm{MWh}$ of total energy to the EV fleet, with a user satisfaction rate of $95.6\%$, ensuring nearly all charging requirements are met. This satisfaction level is within $4\%$ of the oracle MPC (which achieves $99.9\%$) but outperforms all other RL baselines by at least $5\%$ (TD3: $90.2\%$, PPO: $69.4\%$, SAC: $57.7\%$). 
In terms of voltage regulation, PI-TD3 reduces the total number of voltage violations per bus to $104.2$, which is a $20\%$ improvement over TD3 ($127.8$), and $15\%$ lower than PPO ($121.4$). Compared to the oracle MPC, PI-TD3's voltage violation is only $6\%$ higher, indicating near-optimal grid support even without perfect future knowledge. For total voltage violations per step, PI-TD3 achieves $25.3$ violations, $15\%$ lower than TD3 ($30.0$), and only $4\%$ higher than MPC ($24.3$). The average absolute per-unit voltage magnitude violation is also $15\%$ lower for PI-TD3 compared to TD3.

PI-TD3 also maintains competitive charging cost performance, with total costs $19\%$ lower than TD3 and only $23\%$ below MPC. Although CAFAP minimizes user dissatisfaction (charging everyone at full speed), it results in the worst voltage magnitude violations (over $50\%$ higher than PI-TD3) and higher costs. Classic RL methods (SAC, PPO) either sacrifice user satisfaction or grid reliability, as reflected in their lower performance across at least one key metric. Meanwhile, the proposed PI-TD3 can be executed in real-time, requiring only an average $10$ ms per step, while an equivalent MPC-based method would take a few minutes to generate an optimal solution given the scale and the complexity of this MINLP.
Overall, PI-TD3 is the only method that closely matches the oracle MPC in all three objectives (user satisfaction, voltage magnitude regulation, and operational cost), demonstrating the advantage of embedding physical knowledge into the RL training process for large-scale, grid-aware EV charging.

\begin{figure}[t]
  \centering    
      \begin{subfigure}[b]{\linewidth}
    \centering
    \includegraphics[width=0.99\linewidth]{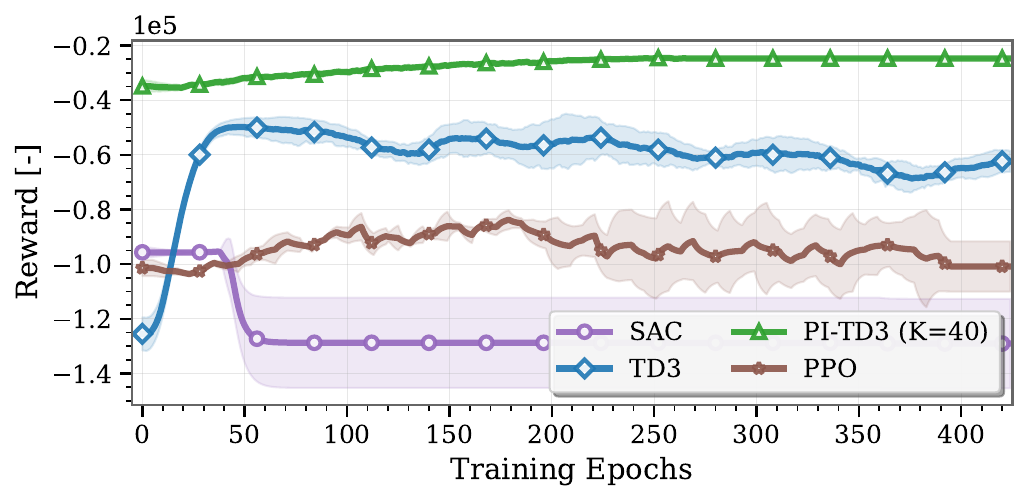}
    \caption{ Convergence curves for PI-TD3 and model-free RL algorithms.}
    \label{fig:sub1RL}
  \end{subfigure}

  \vspace{1em} 

  \begin{subfigure}[b]{\linewidth}
    \centering
    \includegraphics[width=\linewidth]{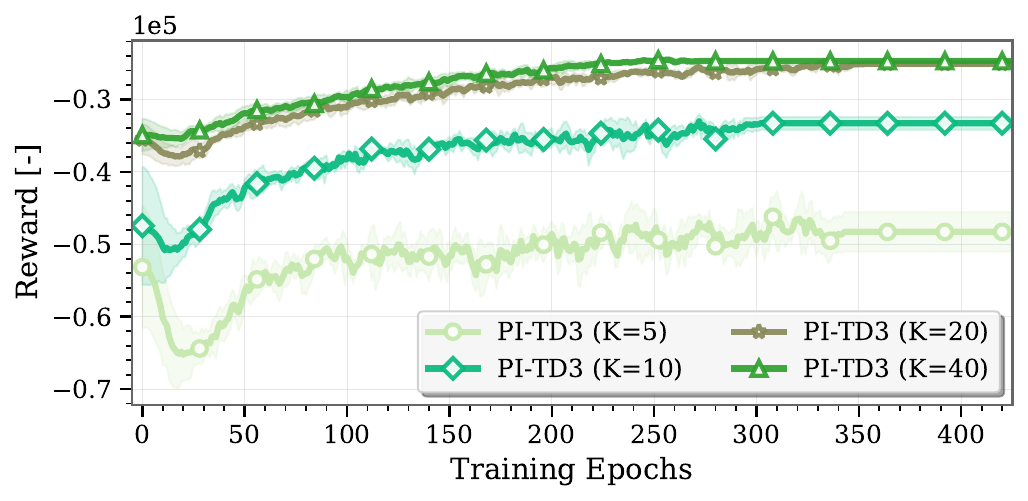}
    \caption{Effect of rollout horizon $K$ on PI-TD3 training, highlighting the existence of an optimal range for $K$ that balances stability and learning speed.}
    \label{fig:sub2RL}
  \end{subfigure}
  
  \caption{Evaluation reward performance comparison during training averaged over five random seeds.}
  
  \label{fig:train_curves}
\end{figure}

\begin{figure}[t]
  \centering

  \begin{subfigure}[b]{\linewidth}
    \centering
    \includegraphics[width=0.99\linewidth]{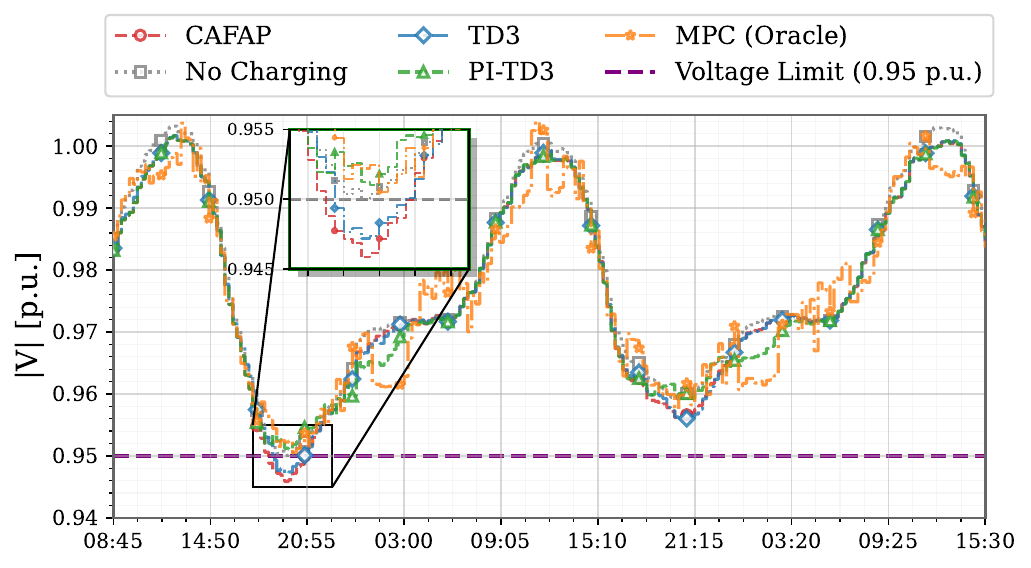}
    \caption{Voltage magnitude at bus 22.}
    \label{fig:sub1}
  \end{subfigure}

  \vspace{1em} 

  \begin{subfigure}[b]{\linewidth}
    \centering
    \includegraphics[width=\linewidth]{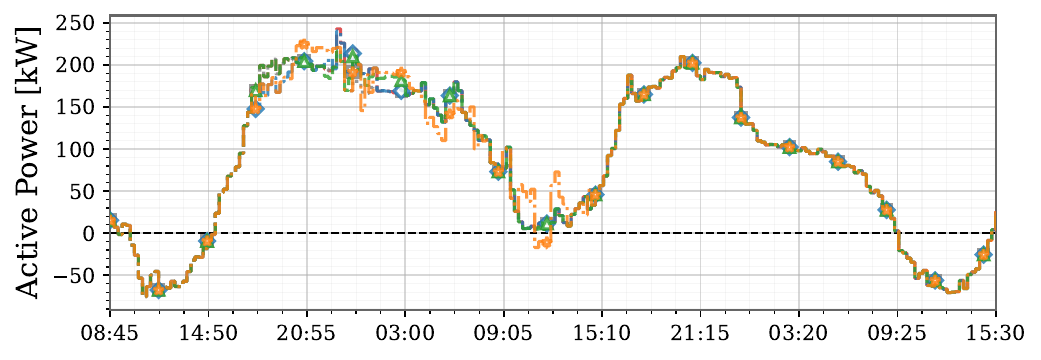}
    \caption{Total active power at bus 22.}
    \label{fig:sub2}
  \end{subfigure}

  \vspace{1em} 

  \begin{subfigure}[b]{\linewidth}
    \centering
    \includegraphics[width=0.99\linewidth]{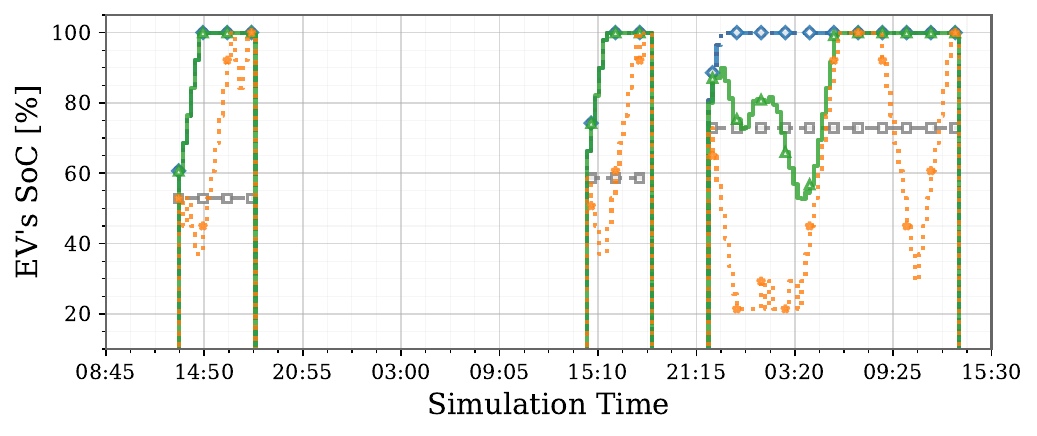}
    \caption{SoC of EVs connected at a charger of bus 22.}
    \label{fig:sub3}
  \end{subfigure}

  \caption{Impact of EV charging algorithm on the bus voltage magnitude and in EV charging. Bus 22 is demonstrated here as a node where applying the proposed algorithm effectively mitigates voltage limit violations.}
  \label{fig:three_subs}
\end{figure}

\subsection{Sample Efficiency and Convergence Analysis}

To evaluate the training efficiency and learning dynamics of PI-TD3, convergence curves and rollout ablation results are compared against state-of-the-art model-free RL baselines.
Figure~\ref{fig:sub1RL} compares the convergence behavior of PI-TD3 and model-free RL algorithms. PI-TD3 rapidly achieves a maximum reward above $-0.3\times10^5$ within the first 75 epochs, whereas model-free TD3 plateaus near $-0.5\times10^5$, and PPO and SAC remain below $-0.8\times10^5$ throughout training. The incorporation of physical knowledge enables PI-TD3 to reach stable, near-optimal policies approximately four times faster than TD3 and with considerably reduced variance across training runs. This proves a marked improvement in sample efficiency and robustness, making PI-TD3 substantially more suitable for large-scale EV charging control where rapid and reliable learning is critical.

The impact of the rollout horizon $K$ on the performance of PI-TD3 is investigated in Figure~\ref{fig:sub2RL}. As $K$ increases, the learning curve improves: PI-TD3 with $K=5$ exhibits noticeably lower final rewards and slower convergence, demonstrating that short rollouts do not provide sufficient gradient information to fully leverage the physics-based environment. Increasing $K$ to $10$ and then $20$ leads to substantial improvements, both in terms of the speed of convergence and the final achieved reward. Notably, the gap between $K=20$ and $K=40$ becomes minimal, with both configurations converging to a similar reward level close to $-0.2\times 10^5$. This plateau indicates that, beyond a certain point, further increasing the rollout horizon offers diminishing returns, as the policy already benefits from sufficiently long, informative trajectories. Moreover, using extremely large $K$ may introduce practical drawbacks, such as higher computational cost and increased risk of accumulating modeling errors or numerical instability over long rollouts. 
Therefore, careful selection of the rollout horizon is crucial for achieving optimal trade-offs between gradient quality, learning efficiency, and computational tractability in PI-RL.

\begin{figure*}[t]
  \centering
  \includegraphics[width=\textwidth]{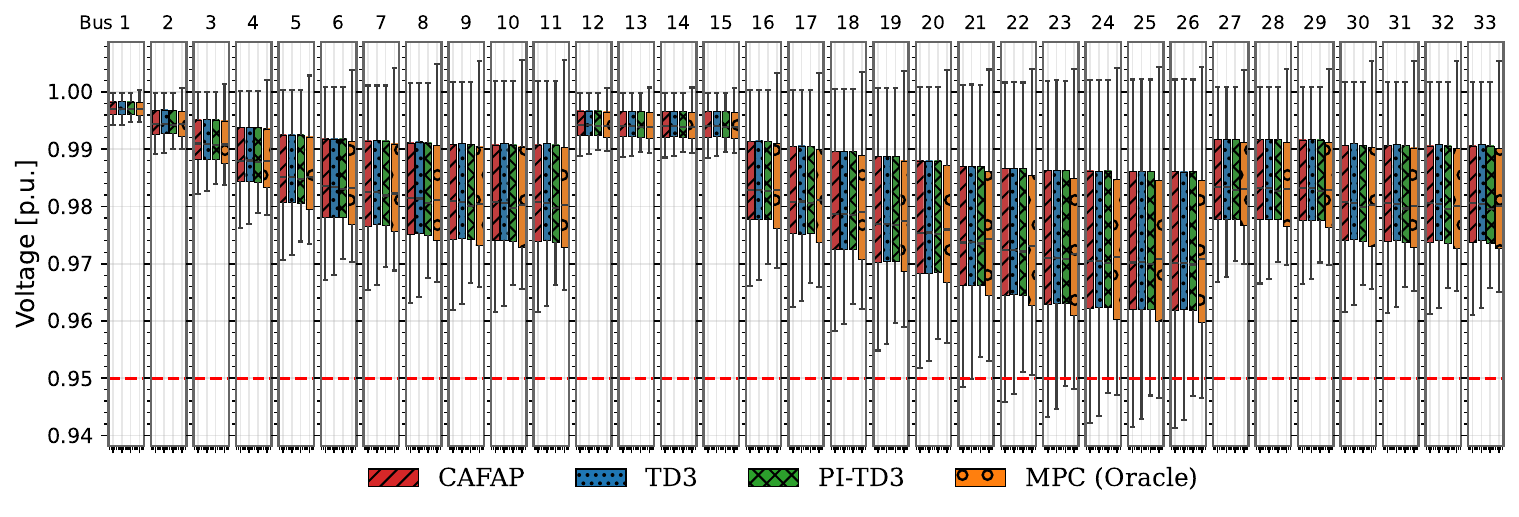} 
  \caption{
Distribution of bus voltage magnitudes for each charging algorithm. Box plots show median, interquartile range, and full voltage range over an experimental evaluation scenario; the red dashed line indicates the operational limit ($0.95$~p.u.).
}
  \label{fig:volt_per_bus}
\end{figure*}

\subsection{Detailed Node-Level and System Voltage Analysis}
To provide a detailed view of how each algorithm operates in a realistic scenario, Figure~\ref{fig:three_subs} examines the voltage and charging profiles at bus $22$ over a representative evaluation day. As shown in Figure~\ref{fig:sub1}, the proposed PI-TD3 algorithm substantially reduces both the frequency and severity of voltage limit violations at this critical bus. In particular, PI-TD3 maintains the voltage magnitude above the $0.95$~p.u. threshold, compared to TD3 and CAFAP baselines that fail to do so.
Figure~\ref{fig:sub2} presents the total active power drawn at bus 22. While all algorithms yield similar aggregate profiles, PI-TD3 selectively modulates the charging load, especially during periods of heightened grid stress, to mitigate voltage violations. This dynamic response highlights the PI-TD3's algorithm’s ability to maintain high charging throughput without sacrificing grid stability.
The charging schedules of individual EVs, as depicted in Figure~\ref{fig:sub3}, demonstrate the diverse and adaptive strategies enabled by PI-TD3. The SoC trajectories reveal that, unlike heuristic or purely model-free baselines, PI-TD3 achieves 100\% user satisfaction at bus 22, ensuring all EVs depart fully charged even under congested conditions. 

System-wide results are summarized in Figure~\ref{fig:volt_per_bus}. Here, the voltage magnitude distributions across all 33 buses (excluding the reference bus) show that PI-TD3 achieves similar median voltages and violation rates as the oracle MPC, despite lacking access to future information. Specifically, PI-TD3 outperforms TD3 by reducing the average per-bus voltage magnitude violations by a noticeable margin, and narrows the gap with the MPC lower bound. However, some violation events persist across all algorithms due to the intentionally overloaded grid design, underscoring the challenging nature of the test scenario.
Overall, these results confirm that PI-TD3 delivers robust, grid-compliant EV charging, achieving an advantageous trade-off between energy delivery, cost savings, and voltage magnitude regulation when compared to state-of-the-art RL and heuristic baselines.

\begin{figure}[t]
  \centering  
  \includegraphics[width=\linewidth]{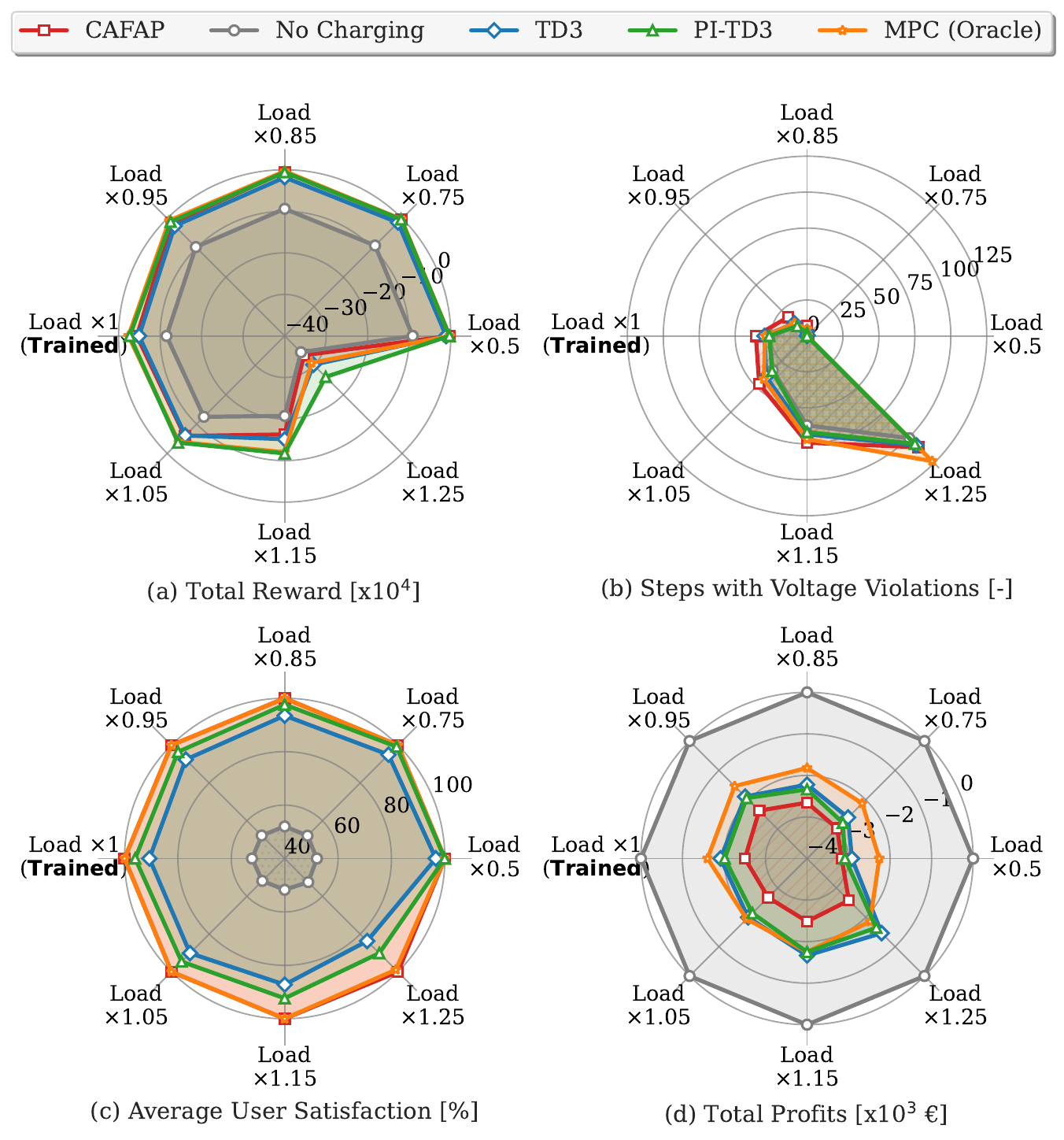} 
  \caption{
Generalization and robustness of charging algorithms to varying grid loading. Each radar plot shows performance: (a) total reward, (b) voltage violation steps, (c) average user satisfaction, and (d) total profits, for load multipliers from $0.5\times$ to $1.5\times$ on the IEEE 34-bus network. The PI-TD3 agent was trained on the reference scenario and evaluated out-of-distribution, demonstrating adaptability.
}
  \label{fig:gen}
\end{figure}

\subsection{Generalization to Unseen Load Profiles}

To assess the robustness of the proposed PI-TD3 algorithm, a generalization study was conducted using modified IEEE 34-bus networks with load scaling factors from $0.5\times$ up to $1.25\times$ nominal demand. The PI-TD3 and classic TD3 agents were exclusively trained on the nominal grid ($1.0\times$ load) and evaluated directly on all other load scenarios, providing an out-of-distribution generalization benchmark.

Figure~\ref{fig:gen} summarizes the performance across four key metrics. As shown in Figure~\ref{fig:gen}a, PI-TD3 maintains top performance in total reward, with values nearly indistinguishable from the oracle MPC in all but the most extreme loads. For grid reliability (Figure~\ref{fig:gen}b), PI-TD3 reduces the number of time steps with voltage magnitude violations compared to standard TD3, particularly as the network becomes more stressed.
In terms of user experience, Figure~\ref{fig:gen}c shows that PI-TD3 keeps average user satisfaction above 90\% across the entire range, a level matched only by MPC and CAFAP, and exceeding the TD3 baseline by 5-15\%  as load increases. Regarding total profits (Figure~\ref{fig:gen}d), PI-TD3 is more cost-efficient than the simple baselines. However, PI-TD3 is also very close to the Oracle when evaluated on cases with higher loads than the one it was trained on.

Notably, PI-TD3 attains these results without retraining or fine-tuning on the new conditions, highlighting its ability to generalize robustly to previously unseen load profiles. By leveraging physical knowledge and differentiable rollouts, PI-TD3 learns policies that remain grid-compliant and economically efficient under a wide spectrum of practical operating conditions, outperforming all model-free and heuristic alternatives in these challenging out-of-distribution tests.

\begin{figure}[t]
  \centering  
  \includegraphics[width=\linewidth]{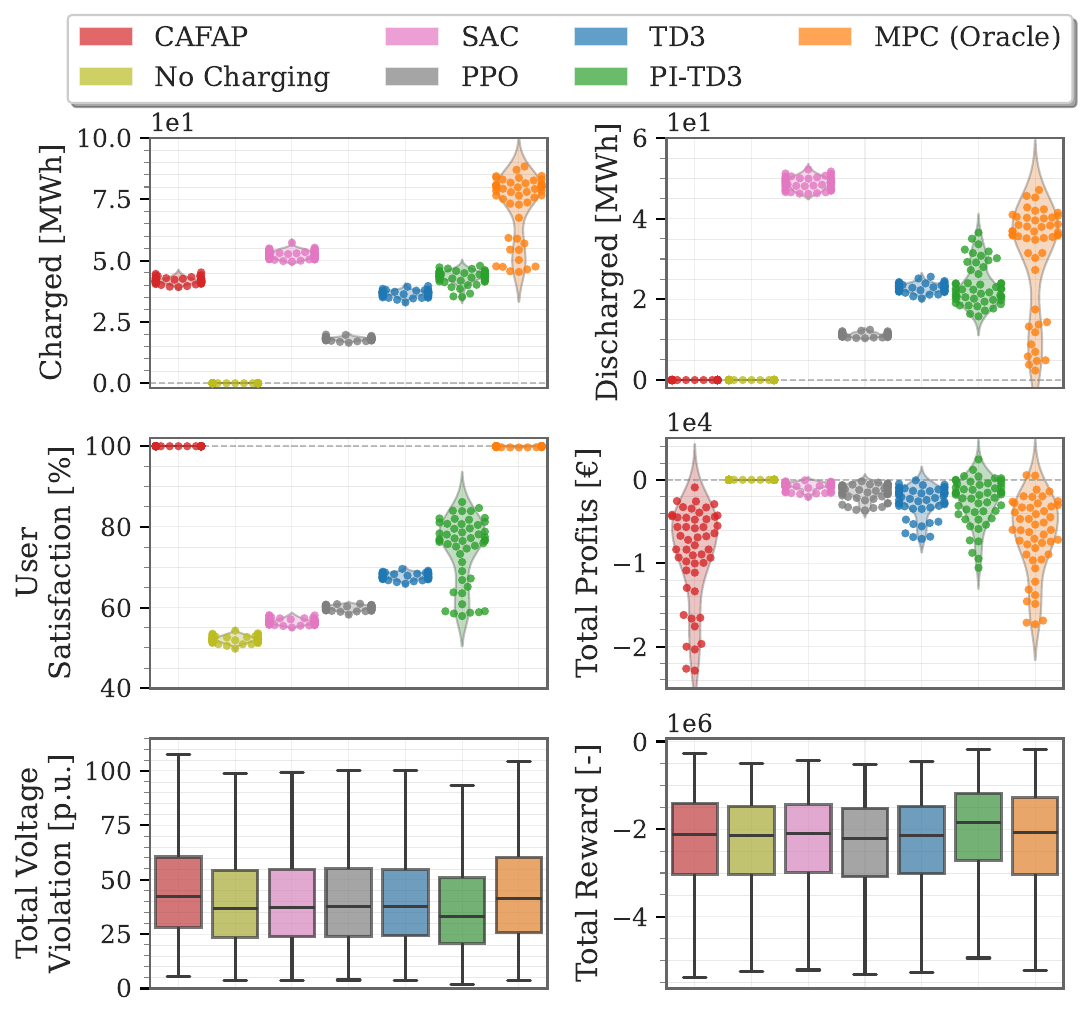} 
  \caption{
Performance comparison of charging algorithms on the IEEE 123-bus network with 500 EVs. Metrics shown include energy charged/discharged, user satisfaction, profits, total voltage violations, and total reward. PI-TD3 matches or surpasses the oracle MPC in most metrics and maintains robust grid operation at scale.
}

  \label{fig:scale}
\end{figure}


\subsection{Scalability Study: Large-Scale Grid with 500 EVs}

To further evaluate the scalability and robustness of the proposed PI-TD3 algorithm, experiments were conducted on the IEEE 123-bus network with 500 distributed EV charging points, a significant increase in both network and EV fleet size. For each algorithm, $50$ independent experimental scenarios were randomly selected. In Figure~\ref{fig:scale}, the first two rows display violin plots, where each dot represents the outcome of a single scenario, thereby visualizing both the overall result distribution and the deviation across trials. These plots reveal that PI-TD3 and the oracle MPC consistently achieve the highest user satisfaction, and the most favorable profits, while classic RL algorithms and heuristics exhibit less diverse performances.
The final row of Figure~\ref{fig:scale}, presents box plots summarizing the total voltage magnitude violations (p.u.) and total rewards. Here, PI-TD3 matches or slightly outperforms MPC in both metrics, achieving median total voltage magnitude violations below $30$ and maintaining a consistently higher total reward than all other methods. The box plots show that the distribution of voltage magnitude violations and rewards for PI-TD3 are notably narrower, indicating reliability and stability in grid operation at this scale.
Overall, these results demonstrate that PI-TD3 not only achieves strong average performance but also maintains low variability and robust grid support even as the problem size and complexity are greatly increased.

\section{Conclusion}
This work introduced PI-TD3, a PI-RL algorithm for large-scale EV smart charging supporting the voltage magnitude of the distribution network.
Pi-TD3 effectively embeds the power flow formulation via differentiable reward signals directly into the learning process, obtaining richer gradient information that accelerates convergence and improves constraint satisfaction.
The proposed PI-TD3 algorithm was evaluated against classic model-free RL approaches, heuristic methods, and an oracle MPC, consistently surpassing all baselines and matching the oracle in voltage magnitude regulation, user satisfaction, and economic performance, even in overloaded and highly variable grid conditions. Extensive experiments on the IEEE 34-bus and 123-bus networks demonstrated the superior generalization, stability, and scalability of PI-TD3, maintaining high performance across hundreds of EVs and diverse scenarios. 
Future research may extend PI-TD3 to additional domains within the smart grid and broader cyber-physical systems, as well as address the integration of non-differentiable dynamics, real-time adaptation, and deployment in real-world pilot studies.

\bibliographystyle{IEEEtran}
\bibliography{ref}

\begin{thebibliography}{10}
\providecommand{\url}[1]{#1}
\csname url@samestyle\endcsname
\providecommand{\newblock}{\relax}
\providecommand{\bibinfo}[2]{#2}
\providecommand{\BIBentrySTDinterwordspacing}{\spaceskip=0pt\relax}
\providecommand{\BIBentryALTinterwordstretchfactor}{4}
\providecommand{\BIBentryALTinterwordspacing}{\spaceskip=\fontdimen2\font plus
\BIBentryALTinterwordstretchfactor\fontdimen3\font minus \fontdimen4\font\relax}
\providecommand{\BIBforeignlanguage}[2]{{%
\expandafter\ifx\csname l@#1\endcsname\relax
\typeout{** WARNING: IEEEtran.bst: No hyphenation pattern has been}%
\typeout{** loaded for the language `#1'. Using the pattern for}%
\typeout{** the default language instead.}%
\else
\language=\csname l@#1\endcsname
\fi
#2}}
\providecommand{\BIBdecl}{\relax}
\BIBdecl

\bibitem{STIASNY2021106696}
J.~Stiasny, T.~Zufferey, G.~Pareschi, D.~Toffanin, G.~Hug, and K.~Boulouchos, ``Sensitivity analysis of electric vehicle impact on low-voltage distribution grids,'' \emph{Electr. Pow. Syst. Res.}, vol. 191, p. 106696, 2021.

\bibitem{HASAN2023101216}
K.~N. Hasan, K.~M. Muttaqi, P.~Borboa, J.~Scira, Z.~Zhang, and M.~Leishman, ``Distribution network voltage analysis with data-driven electric vehicle load profiles,'' \emph{Sustainable Energy, Grids and Networks}, vol.~36, p. 101216, 2023.

\bibitem{mattos2024}
M.~M. Mattos, J.~A.~G. Archetti, L.~d.~A. Bitencourt, A.~Wallberg, V.~Castellucci, B.~H. Dias, and J.~G. de~Oliveira, ``Analysis of voltage control using v2g technology to support low voltage distribution networks,'' \emph{IET Generation, Transmission \& Distribution}, vol.~18, no.~6, pp. 1133--1157, 2024.

\bibitem{KNEZOVIC2016274}
K.~Knezović and M.~Marinelli, ``Phase-wise enhanced voltage support from electric vehicles in a danish low-voltage distribution grid,'' \emph{Electr. Pow. Syst. Res.}, vol. 140, pp. 274--283, 2016.

\bibitem{M20251095}
K.~M., M.~D., and K.~Rajagopal, ``Enhancing voltage control and regulation in smart micro-grids through deep learning - optimized ev reactive power management,'' \emph{En. Rep.}, vol.~13, pp. 1095--1107, 2025.

\bibitem{ELALFY2025100872}
D.~A. Elalfy, E.~Gouda, M.~F. Kotb, V.~Bureš, and B.~E. Sedhom, ``Frequency and voltage regulation enhancement for microgrids with electric vehicles based on red panda optimizer,'' \emph{Energy Conversion and Management: X}, vol.~25, p. 100872, 2025.

\bibitem{wevj16060292}
B.~Khan, Z.~Ullah, and G.~Gruosso, ``Enhancing grid stability through physics-informed machine learning integrated-model predictive control for electric vehicle disturbance management,'' \emph{World Electric Vehicle Journal}, vol.~16, no.~6, 2025.

\bibitem{10138878}
S.~Ke, J.~Yang, L.~Chen, P.~Fan, X.~Shi, G.~Li, and F.~Wu, ``A frequency control strategy for ev stations based on mpc-vsg in islanded microgrids,'' \emph{IEEE Trans. on Ind. Inf.}, vol.~20, no.~2, pp. 1819--1831, 2024.

\bibitem{SINGH2023100972}
S.~Singh and M.~Verma, ``Smart charging schedule of plug-in electric vehicles for voltage support: A prosumer-centric approach,'' \emph{Sustainable Energy, Grids and Networks}, vol.~33, p. 100972, 2023.

\bibitem{SuttonReinforcementIntroduction}
R.~S. Sutton and A.~G. Barto, \emph{Reinforcement learning : an introduction}.\hskip 1em plus 0.5em minus 0.4em\relax Bradford Books, 2018.

\bibitem{8892476}
Q.~Yang, G.~Wang, A.~Sadeghi, G.~B. Giannakis, and J.~Sun, ``Two-timescale voltage control in distribution grids using deep reinforcement learning,'' \emph{IEEE Trans. on Smart Grid}, vol.~11, no.~3, pp. 2313--2323, 2020.

\bibitem{liu2023DRL}
D.~Liu, P.~Zeng, S.~Cui, and C.~Song, ``Deep reinforcement learning for charging scheduling of electric vehicles considering distribution network voltage stability,'' \emph{Sensors}, vol.~23, no.~3, p. 1618, 2023.

\bibitem{SHIBL2023494}
M.~M. Shibl, L.~S. Ismail, and A.~M. Massoud, ``Electric vehicles charging management using deep reinforcement learning considering vehicle-to-grid operation and battery degradation,'' \emph{En. Rep.}, vol.~10, pp. 494--509, 2023.

\bibitem{fan2024safeRL}
J.~Fan, A.~Liebman, and H.~Wang, ``Safety-aware reinforcement learning for electric vehicle charging station management in distribution network,'' in \emph{2024 IEEE Power \& Energy Society General Meeting (PESGM)}, 2024, pp. 1--5.

\bibitem{hossain2024efficient}
R.~R. Hossain, T.~Yin, Y.~Du, D.~Bienstock, and G.~Zussman, ``Efficient learning of power grid voltage control strategies via model-based deep reinforcement learning,'' \emph{Machine Learning}, vol. 113, pp. 2675--2700, 2024.

\bibitem{9805763}
D.~Hu, Z.~Ye, Y.~Gao, Z.~Ye, Y.~Peng, and N.~Yu, ``Multi-agent deep reinforcement learning for voltage control with coordinated active and reactive power optimization,'' \emph{IEEE Trans. on Smart Grid}, vol.~13, no.~6, pp. 4873--4886, 2022.

\bibitem{9756505}
S.~Li, W.~Hu, D.~Cao, Z.~Zhang, Q.~Huang, Z.~Chen, and F.~Blaabjerg, ``Ev charging strategy considering transformer lifetime via evolutionary curriculum learning-based multiagent deep reinforcement learning,'' \emph{IEEE Trans. on Smart Grid}, vol.~13, no.~4, pp. 2774--2787, 2022.

\bibitem{dulac2021challenges}
G.~Dulac-Arnold, N.~Levine, D.~J. Mankowitz, J.~Li, C.~Paduraru, S.~Gowal, and T.~Hester, ``Challenges of real-world reinforcement learning: definitions, benchmarks and analysis,'' \emph{Machine Learning}, vol. 110, pp. 2419--2468, 2021.

\bibitem{karniadakis2021physics}
G.~E. Karniadakis, I.~G. Kevrekidis, L.~Lu, P.~Perdikaris, S.~Wang, and L.~Yang, ``Physics-informed machine learning,'' \emph{Nature Reviews Physics}, vol.~3, pp. 422--440, 2021.

\bibitem{10905642}
N.~F. Kamal, A.~Sharida, S.~Bayhan, H.~Abu-Rub, and H.~Alnuweiri, ``Enhancing electric vehicle charging predictions: A physics-informed neural network approach,'' in \emph{IECON 2024 - 50th Annual Conference of the IEEE Industrial Electronics Society}, 2024, pp. 1--6.

\bibitem{lim2024evpin}
H.~Lim, J.~W. Lee, J.~Boyack, and J.~B. Choi, ``Ev-pinn: A physics-informed neural network for predicting electric vehicle dynamics,'' 2024.

\bibitem{kaseb2025adaptiveinformeddeepneural}
Z.~Kaseb, S.~Orfanoudakis, P.~P. Vergara, and P.~Palensky, ``Adaptive informed deep neural networks for power flow analysis,'' 2025.

\bibitem{KUANG2024123059}
H.~Kuang, H.~Qu, K.~Deng, and J.~Li, ``A physics-informed graph learning approach for citywide electric vehicle charging demand prediction and pricing,'' \emph{Applied Energy}, vol. 363, p. 123059, 2024.

\bibitem{10113230}
D.~Cao, J.~Zhao, J.~Hu, Y.~Pei, Q.~Huang, Z.~Chen, and W.~Hu, ``Physics-informed graphical representation-enabled deep reinforcement learning for robust distribution system voltage control,'' \emph{IEEE Trans. on Smart Grid}, vol.~15, no.~1, pp. 233--246, 2024.

\bibitem{9916278}
J.~Gao, S.~Chen, X.~Li, and J.~Zhang, ``Transient voltage control based on physics-informed reinforcement learning,'' \emph{IEEE Journal of Radio Frequency Identification}, vol.~6, pp. 905--910, 2022.

\bibitem{ZHANG2024109641}
B.~Zhang, D.~Cao, W.~Hu, A.~M. Ghias, and Z.~Chen, ``Physics-informed multi-agent deep reinforcement learning enabled distributed voltage control for active distribution network using pv inverters,'' \emph{Int. Journal of Electr. Power \& En. Syst.}, vol. 155, p. 109641, 2024.

\bibitem{10418941}
A.~Biswas, M.~Acquarone, H.~Wang, F.~Miretti, D.~A. Misul, and A.~Emadi, ``Safe reinforcement learning for energy management of electrified vehicle with novel physics-informed exploration strategy,'' \emph{IEEE Trans. on Transp. Electr.}, vol.~10, no.~4, pp. 9814--9828, 2024.

\bibitem{orfanoudakis2024ev2gym}
S.~Orfanoudakis, C.~Diaz-Londono, Y.~Emre~Yılmaz, P.~Palensky, and P.~P. Vergara, ``Ev2gym: A flexible v2g simulator for ev smart charging research and benchmarking,'' \emph{IEEE Transactions on Intelligent Transportation Systems}, vol.~26, no.~2, p. 2410–2421, Feb. 2025.

\bibitem{OCPP2.1}
``Open charge point protocol (ocpp) 2.1, edition 1,'' Open Charge Alliance, Technical Report \& Protocol Specification, Jan. 2025.

\bibitem{GIRALDO2022108326}
J.~S. Giraldo, O.~D. Montoya, P.~P. Vergara, and F.~Milano, ``A fixed-point current injection power flow for electric distribution systems using laurent series,'' \emph{Electr. Pow. Syst. Res.}, vol. 211, p. 108326, 2022.

\bibitem{xing2024stabilizing}
E.~Xing, V.~Luk, and J.~Oh, ``Stabilizing reinforcement learning in differentiable multiphysics simulation,'' 2024.

\bibitem{pmlr-v80-fujimoto18a}
S.~Fujimoto, H.~van Hoof, and D.~Meger, ``Addressing function approximation error in actor-critic methods,'' in \emph{Proceedings of the 35th ICML}, vol.~80.\hskip 1em plus 0.5em minus 0.4em\relax PMLR, 10--15 Jul 2018, pp. 1587--1596.

\bibitem{Orfanoudakis2024}
S.~Orfanoudakis, V.~Robu, E.~M. Salazar, P.~Palensky, and P.~P. Vergara, ``Scalable reinforcement learning for large-scale coordination of electric vehicles using graph neural networks,'' \emph{Communications Engineering}, vol.~4, no.~1, p. 118, 2025.

\bibitem{HOU2025100457}
S.~Hou, S.~Gao, W.~Xia, E.~M. {Salazar Duque}, P.~Palensky, and P.~P. Vergara, ``Rl-adn: A high-performance deep reinforcement learning environment for optimal energy storage systems dispatch in active distribution networks,'' \emph{Energy and AI}, vol.~19, p. 100457, 2025.

\bibitem{DHPC2024}
{D}elft {H}igh {P}erformance {C}omputing~{C}entre ({DHPC}), ``{D}elft{B}lue {S}upercomputer ({P}hase 2),'' 2024.

\end{thebibliography}

\end{document}